\documentclass[10pt,notitlepage,aps,prl,twocolumn]{revtex4-1}
\usepackage{lmodern}
\usepackage{textcomp}
\usepackage{graphicx}
\usepackage{amsmath}
\usepackage{graphicx}
\usepackage{bm}
\usepackage{array}
\usepackage{dcolumn}
\usepackage{hepparticles}
\usepackage{heppennames}
\usepackage{hepnicenames}
\usepackage{hyperref}
\usepackage{lineno}

\renewcommand{\figurename}{\textbf{Figure}}
\renewcommand{\thefigure}{\textbf{\arabic{figure}}} 

\newcommand{\FigureOne}{
\begin{figure}[h!]
\includegraphics[]{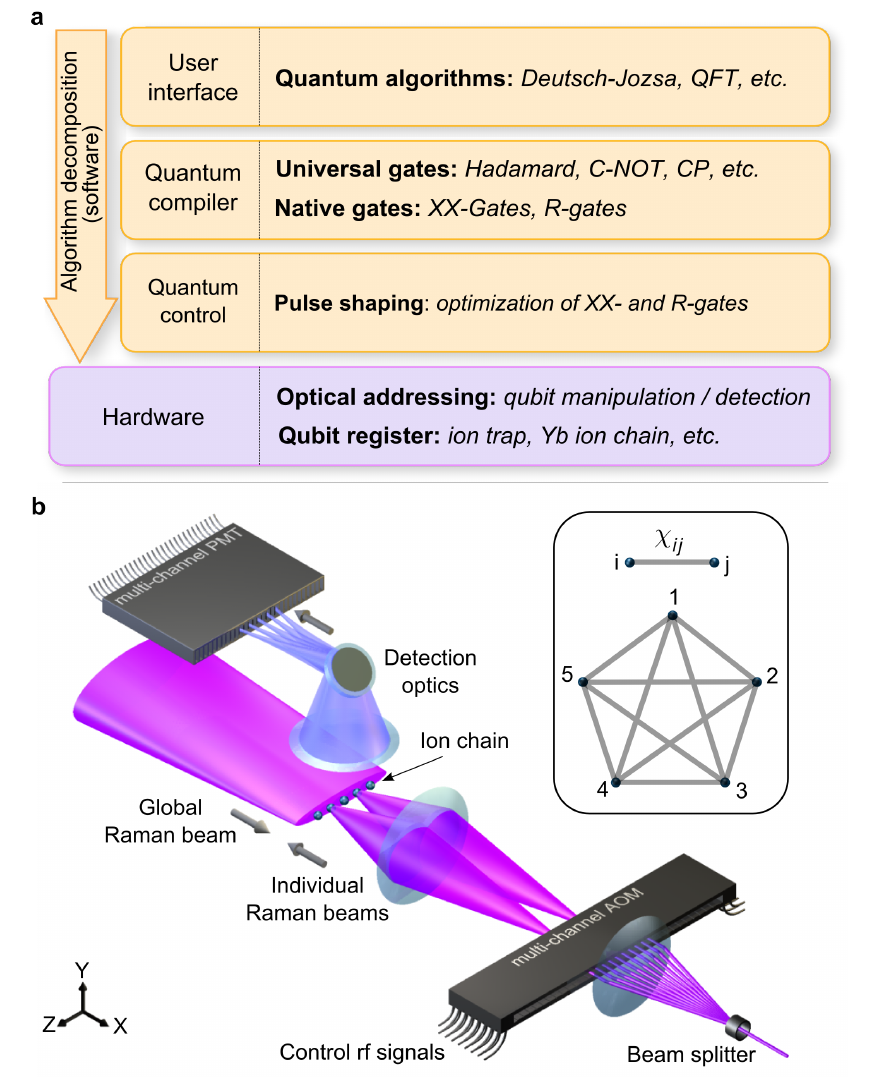}
\caption{\textbf{Computation architecture.} \textbf{(a)} Decomposition of algorithms from the user interface and software operations to the physical hardware. \textbf{(b)} Hardware setup. A linear chain of trapped ion qubits along the Z-axis is shown at the center of the figure. An imaging objective collects ion fluorescence along the Y-axis and maps each ion onto a multichannel photo-multiplier tube (PMT) for measurement of individual qubits. Counterpropagating Raman beams along the X-axis perform qubit operations. A diffractive beam splitter creates an array of static Raman beams that are individually switched using a multi-channel acousto-optic modulator (AOM) to perform qubit-selective gates. By modulating appropriate addressing beams, any single-qubit rotation or two-qubit Ising (XX) gate can be realized. For the two-qubit gates between qubits $i$ and $j$, we can continuously tune the nonlinear gate angle $\chi_{ij}$. This represents a system of qubits with fully connected and reconfigurable spin-spin Ising interactions (inset).
}
\label{fig:Setup}
\end{figure}}

\newcommand{\FigureTwo}{
\begin{figure}[t]
\includegraphics[]{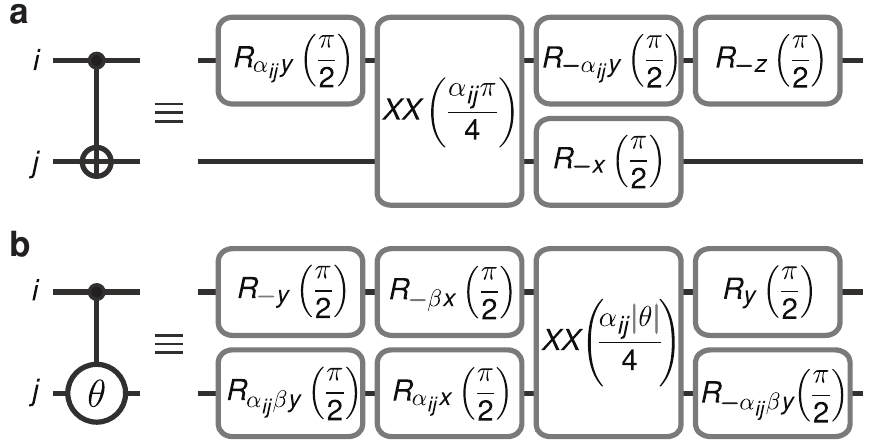}
\caption{
\textbf{Two-qubit modular gates.} \textbf{(a)} Decomposition of the CNOT gate. The geometric phase $\chi_{ij}$ of the XX-gate is $\pm \pi/4$, and we define $\alpha_{ij}=\text{sgn}(\chi_{ij})$. \textbf{(b)} Decomposition of the controlled-phase gate where $\beta =\text{sgn}(\theta)$ for the controlled phase $\theta$. The geometric phase of the XX-gate is adjusted such that $\chi_{ij}=\alpha_{ij}|\theta|/4$.
}
\label{fig:ModularGates}
\end{figure}}

\newcommand{\FigureThree}{
\begin{figure*}[t]
\centering
\includegraphics[]{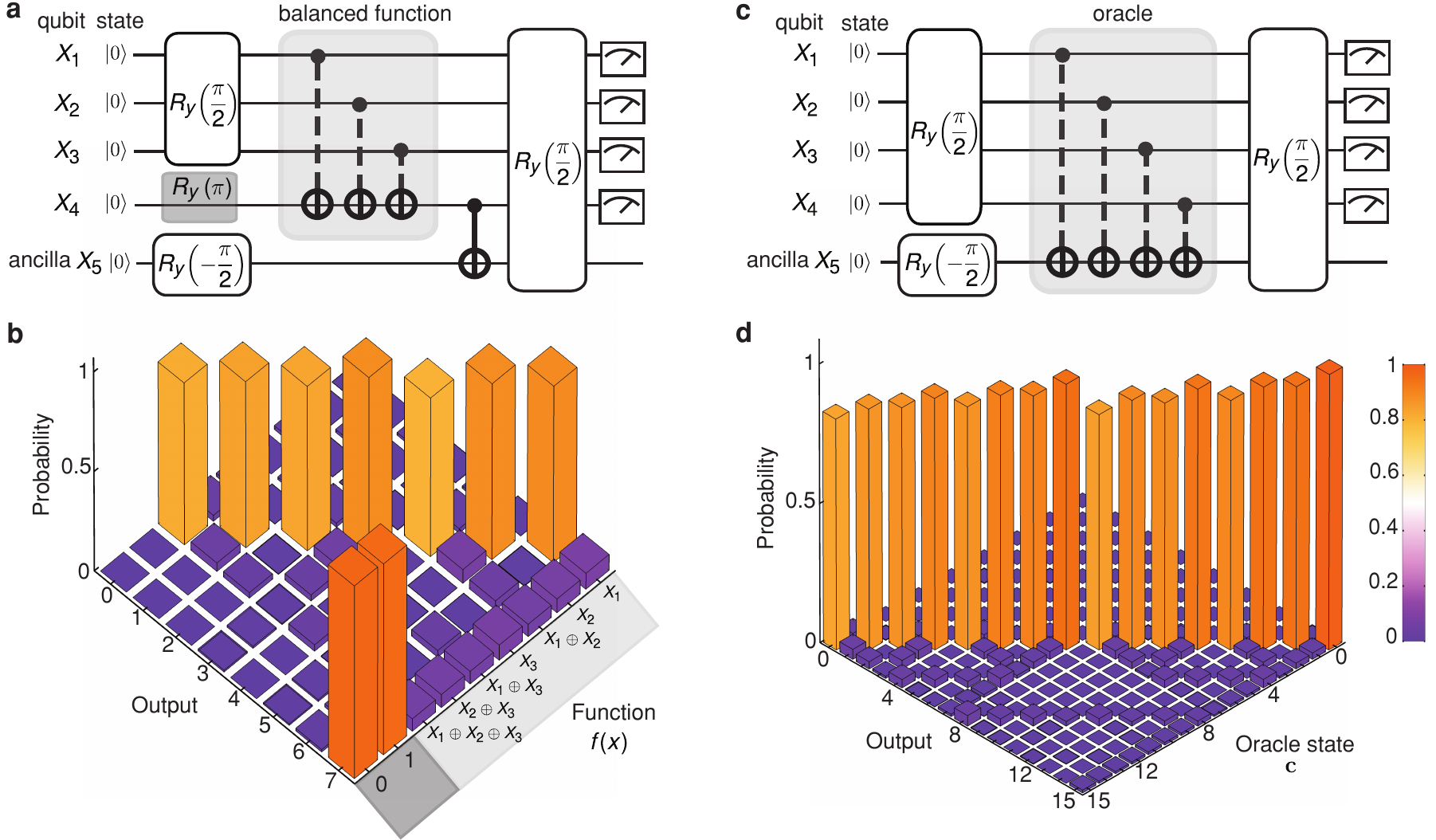}
\caption{\textbf{Quantum Algorithms.} \textbf{(a)} The Deutsch-Jozsa algorithm circuit on 5 ions. The oracle is implemented through gates shown in the shaded regions of the circuit. For balanced function oracles we apply each of the seven possible CNOT combinations, indicated in light gray. For the constant functions, we prepare $X_4=0$ or $1$ as indicated in dark gray. \textbf{(b)} Measured populations of the output state for various functions, conditioned upon measuring $X_4=1$. The two constant functions $f=0$ and $f=1$ are indicated in dark gray, and the seven balanced functions given by particular CNOT gate combinations are indicated in light gray. Measurement of the output $\{X_1 X_2 X_3\}=\{111\}=7$  indicates a constant function, while any other value ($0-6$) indicates a balanced function. \textbf{(c)} The Bernstein-Vazirani algorithm circuit. The shaded region contains programmed CNOT gate combinations used to implement different oracle states $\textbf{c}$. \textbf{(d)} Measured output population for various oracle states. The output is the inverted oracle state $\bar{\textbf{c}}$. Data represented in \textbf{b, d} are obtained by sampling over $\sim$20,000 experimental repetitions for each function or the oracle state \textbf{c} and the errors for the success probabilities in each case are statistical estimates.
}
\label{fig:DJBV}
\end{figure*}}

\newcommand{\FigureFour}{
\begin{figure*}[t]
\centering
\includegraphics[]{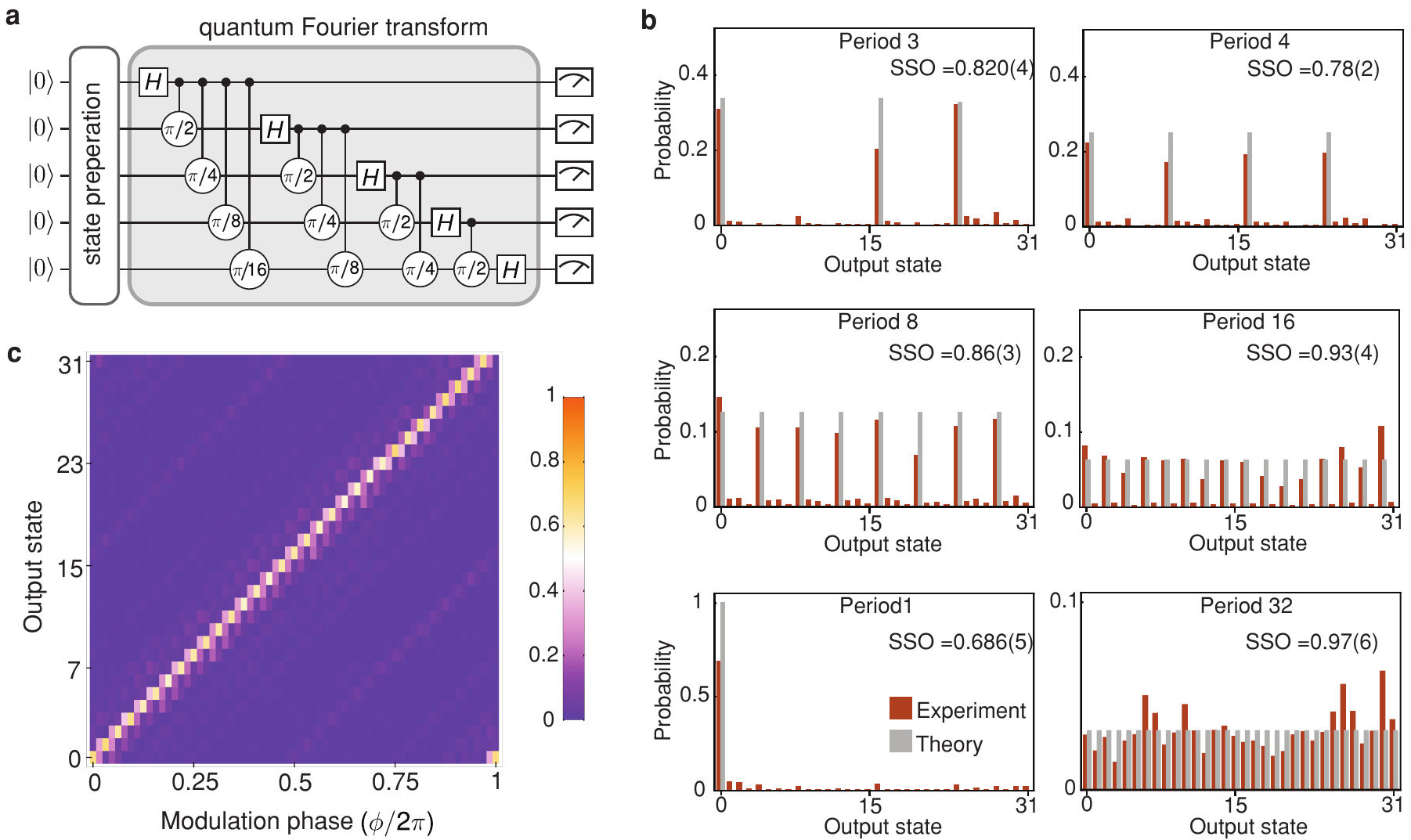}
\caption{\textbf{Quantum Fourier transform protocol.} \textbf{(a)} Experimental sequence for implementation and verification of the QFT. `State preparation' consists of single qubit rotations that create a phase and amplitude modulation of the coefficients $\{C_k\}$ of the input state $\sum_{k=0}^{31} C_k \ket{k}$. The shaded gray region contains a sequence of modular gates for implementing the QFT, which is then followed by a measurement of the register. \textbf{(b)} Quantum period finding. Input states are prepared using single-qubit rotations to modulate the 32 state amplitudes with periods 1, 3, 4, 8, 16, and 32 (see methods). The squared statistical overlap (SSO) \cite{Chiaverini05} signifies the fidelity of the protocol where the error is a statistical estimate over 8,000 experimental repetitions. \textbf{(c)} Quantum phase estimation using five measurement qubits. The plot shows populations in the output state that estimates the given phase modulation $\phi$ of the input state amplitudes $\{C_k\}$. We observe the correct value of the phase in each case with a probability $> 0.6$. The experiment is repeated 8,000 times for each value of $\phi$.
}
\label{fig:QFT}
\end{figure*}}

\begin{document}
\newcommand{\ket}[1]{|#1\rangle}
\newcommand{\bra}[1]{\langle #1|}
\newcommand{\Yb}{$^{171}{\rm{Yb}}^{+} $}
\newcommand{\up}{\uparrow}
\newcommand{\dn}{\downarrow}
\newcommand{\upr}{\uparrow\rangle}
\newcommand{\dnr}{\downarrow\rangle}
\newcommand{\sx}{\hat{\sigma}_{x}}
\newcommand{\sy}{\hat{\sigma}_{y}}
\newcommand{\sz}{\hat{\sigma}_{z}}

\title{Demonstration of a small programmable quantum computer with atomic qubits}

\author{S. Debnath}
\author{N. M. Linke}
\author{C. Figgatt}
\author{K. A. Landsman}
\author{K. Wright}
\author{C. Monroe}
\affiliation{Joint Quantum Institute, Department of Physics, and Joint Center for Quantum Information and Computer Science, University of Maryland, College Park, MD  20742.}


\maketitle

\textbf{Quantum computers can solve certain problems more efficiently than any possible conventional computer. Small quantum algorithms have been demonstrated on multiple quantum computing platforms, many specifically tailored in hardware to implement a particular algorithm or execute a limited number of computational paths \cite{Linden98, Vandersypen01, Gulde03, Shi10, Brainis03, Chiaverini05, Brickman05, Dicarlo09, Vandersypen01, Lopez12, Monz15}. Here, we demonstrate a five-qubit trapped-ion quantum computer that can be programmed in software to implement arbitrary quantum algorithms by executing any sequence of universal quantum logic gates. We compile algorithms into a fully-connected set of gate operations that are native to the hardware and have a mean fidelity of $98\:\%$. Reconfiguring these gate sequences provides the flexibility to implement a variety of algorithms without altering the hardware. As examples, we implement the Deutsch-Jozsa (DJ) \cite{Deutsch92} and Bernstein-Vazirani (BV) \cite{Bernstein97} algorithms with average success rates of $95\:\%$ and $90\:\%$, respectively. We also perform a coherent quantum Fourier transform (QFT) \cite{Shor97,Nielsen02} on five trapped-ion qubits for phase estimation and period finding with average fidelities of $62\:\%$ and $84\:\%$, respectively. This small quantum computer can be scaled to larger numbers of qubits within a single register, and can be further expanded by connecting several such modules through ion shuttling \cite{Kielpinski02} or photonic quantum channels \cite{Monroe14}.}
\FigureOne

Implementing a scalable programmable quantum computing architecture requires high fidelity initialization and detection at the individual qubit level and pristine control of interactions between qubits. While most physical platforms have nearest-neighbor interactions only, a multi-qubit trapped-ion system features an intrinsic long-range interaction that is optically gated and connects any pair of qubits \cite{Cirac95, Molmer99}. Unlike solid-state implementations \cite{Barends14, Hill15}, the quantum circuitry is determined by external fields, and hence can be programmed and reconfigured without altering the structure of the qubits themselves. By optically resolving individual ions, we implement single-qubit rotations and arbitrary two-qubit gates by directly addressing pairs of ions without additional overhead such as moving information through local couplings \cite{Barends14, Gottesman00} or hiding qubit populations in additional auxiliary states \cite{Monz15}. Such native gates can then be used to construct modular logic gates that can be called in reconfigurable algorithm sequences. We observe a mean fidelity of $98\:\%$ in these native operations without the use of spin echo or dynamical decoupling techniques \cite{Green15,Monz15,Ballance16}. This bottom-up approach can be adapted for large scale computation using micro-fabricated ion traps with integrated optics \cite{Merrill11} and high optical access, and we expect gate fidelities can exceed $99.9\:\%$ with straightforward improvements on the classical control \cite{Ballance16, Gaebler16}.

The programmable and reconfigurable nature of the ion trap quantum computer is illustrated by a hierarchy of operations from software to hardware, shown in figure \ref{fig:Setup}a. At the top is a high-level user interface that specifies the desired algorithm, represented by a standard family of modular universal logic gates such as Hadamard (H), controlled-NOT (CNOT), and controlled-phase (CP) gates \cite{Nielsen02}. Next, a quantum compiler translates the universal gates into gates native to the hardware, which in our case are two-qubit Ising (XX) gates \cite{Molmer99} and single-qubit rotation (R) gates \cite{Nielsen02}. Finally, these native gates are decomposed into laser pulses that are pre-calculated to effect the desired qubit operation through the Coulomb-coupled motion while disentangling the motion at the end of the gates \cite{Choi14}.

At the hardware level, the processor consists of trapped \Yb atomic ion qubits with information stored in the hyperfine ``clock'' states $\ket{0}\equiv |F=0;m_{F}=0 \rangle$ and $\ket{1}\equiv |F=1;m_{F}=0 \rangle$ of the $^{2}S_{1/2}$ electronic ground level with a qubit frequency splitting of $\nu_0=12.642821\:$GHz \cite{Olmschenk07}. Here, $F$ and $m_F$ denote the quantum numbers associated with the total atomic angular momentum and its projection along the quantization axis defined by an applied magnetic field of $5.2\:$G. We measure a qubit coherence time in excess of $0.5\:$s, and with magnetic shielding we expect this to improve to be longer than $1000\:$s \cite{Fisk97}.

We confine the ions in a linear radio frequency (rf) Paul trap, with radial and axial trap frequencies $\nu_x = 3.07\:$MHz and $\nu_z=0.27\:$MHz, respectively. The ions are laser-cooled to near their motional ground state and form a linear crystal with a spacing of $\sim 5\:\mu$m for $n=5$ ions. A computation is performed by first initializing all qubits to state $\ket{0}$ through optical pumping \cite{Olmschenk07}. This is followed by quantum gates, implemented by a series of coherent rotations using stimulated Raman transitions driven by a $355\:$nm mode-locked laser, where the beat-note between two counterpropagating Raman beams drives qubit and motional transitions \cite{Hayes10}. To achieve individual addressing, we split one of the Raman beams into a static array of beams, each of which is directed through an individual channel of a multi-channel acousto-optic modulator (AOM) and focused onto its respective ion, as shown in figure \ref{fig:Setup}b. Finally, the qubit register is measured with high fidelity (see methods) by driving the $^{2}S_{1/2}\rightarrow ^{2}P_{1/2}$ cycling transition near $369\:$nm and simultaneously collecting the resulting state-dependent fluorescence from each ion using high-resolution optics and a multi-channel photo-multiplier tube (PMT).

\FigureTwo

The lowest level of qubit control consists of native single- and two-qubit operations. We perform single-qubit rotations $R_{\phi}(\theta)$ by tuning the Raman beat-note to qubit resonance $\nu_0$. Here, the rotation angle $\theta$ and axis $\phi$ are determined by the duration and phase offset of the beat-note, which is programmed through rf signals on appropriate AOM channels. Two-qubit XX-gates are performed by invoking an effective spin-spin Ising interaction between qubits mediated by the collective modes of motion of the chain \cite{Molmer99}. Here, we apply Raman beat-notes tuned close to $\nu_0 \pm \nu_x$ that coherently couple the spins to all modes of motion. A pulse shaping technique \cite{Choi14}  disentangles the motion at the end of the gate, resulting in a two-qubit entangling rotation of any amount $XX(\chi_{i,j})$. Here, the geometric phase $\chi_{i,j}$ originates from the integrated Ising interaction \cite{Molmer99,Choi14}, the sign $\alpha_{i,j}=\pm 1$ of which arises from the Coulomb interaction between qubits $i$ and $j$ (Fig. \ref{fig:Setup}b inset). We pre-calculate and optimize XX-gate pulse shapes off-line for all $\{i,j\}$ to achieve high fidelity while keeping the gates relatively fast (see methods). 

We use these native R and XX-gates to construct standard logic gates, which can be called by a quantum algorithm. For instance, we implement the single-qubit Hadamard gate as $H=R_x(-\pi) R_y(\pi/2)$ and the Z-rotation as $R_z(\theta)=R_y(-\pi/2) R_x(\theta) R_y(\pi/2)$. Two-qubit logic gates such as CP and CNOT are compiled to account for the signs of the CP rotation angle $\beta$ and the Ising interaction $\alpha_{i,j}$, making them independent of $\{i,j\}$ and therefore modular (Fig. \ref{fig:ModularGates}). At the highest level we program arbitrary sequences of such logic gates as required to implement any quantum algorithm.

\FigureThree

We first implement the Deutsch-Jozsa algorithm \cite{Deutsch92}, which determines whether a given function (the ``oracle'') is constant or balanced. A function that has an n-bit input and a 1-bit output ($f:\{0,1,2,...,2^n-1\}\rightarrow \{0,1\}$) is balanced when exactly half of the inputs result in the output $0$ and the other half in the output $1$, while a constant function assumes a single value irrespective of the input. In our setup we program $7$ out of the $70$ possible oracles of three-qubit balanced functions by using seven different sequences of CNOT gates between each of the three qubits in the control register $x=\{X_1 X_2 X_3\}$ and the function register $X_4$ (Fig. \ref{fig:DJBV}a). We program the two constant functions by setting $X_4$ to either $0$ or $1$. Executing the algorithm starts with preparing the control register in the superposition state $\ket{x}=\frac{1}{\sqrt{8}}\sum_{k=0}^7\ket{k}$, followed by the function evaluation oracle. A CNOT is then performed between the function register $X_4$ and the ancilla qubit $X_5$ (initially set to $\frac{1}{\sqrt{2}}(\ket{0}-{\ket{1}})$). All qubits are then rotated and measured (except for the ancilla) as shown in figure \ref{fig:DJBV}a. Finally, a measurement of $x$ (conditioned upon $X_4 = 1$, occurring with $50\%$ probability) determines if the function is constant or balanced. Measurement of the output $x=\{111\}$  indicates a constant function, while any other value indicates a balanced function (see supplementary materials). The average success probability is $0.967(2)$ for constant and $0.932(3)$ for balanced functions (Fig. \ref{fig:DJBV}b).

\FigureFour

The Bernstein-Vazirani algorithm is a variant of the DJ algorithm where the oracle function is an inner product of two n-bit strings: $f_{\textbf{c}}(\textbf{x}) = \textbf{c} \cdot \textbf{x}$. Here, the aim is to determine the vector $\textbf{c}=\{c_1 c_2 ... c_n\}$ in a single trial \cite{Bernstein97}. We program all 16 instances of the four-bit oracle that evaluate the function $f_{\textbf{c}}(\textbf{x}) \oplus X_5$. This is achieved by applying a particular pattern of CNOT gates, determined by \textbf{c}, between $\textbf{x}=\{X_1 X_2 X_3 X_4\}$ and $X_5=\frac{1}{\sqrt{2}}(\ket{0}-\ket{1})$ (Fig \ref{fig:DJBV}c). For example, if $\textbf{c}=\{0101\}$ then CNOT gates are applied between $X_2,X_5$ and $X_4,X_5$. We start with a superposition state $\ket{x}=\frac{1}{\sqrt{16}}\sum_{k=0}^{15}\ket{k}$, followed by the oracle. Finally, applying a global $R_y(\pi/2)$ rotation produces the output state $\bar{\textbf{c}}$, which is the inverse of $\textbf{c}$. In the experiment, a single-shot measurement of the correct outcome $\bar{\textbf{c}}$ is obtained with a probability of $0.903(2)$ (Fig \ref{fig:DJBV}d), averaged over all possible oracle states.

Exponential speed-up of many quantum algorithms arises from the fact that parallel function evaluation is performed on a superposition of all classical input states of an n-bit string. These evaulation paths are then interfered using a quantum Fourier transform (QFT) to produce the desired solution \cite{Nielsen02}. One such example is the order-finding protocol in Shor's quantum factorization algorithm \cite{Shor97}. Another application is solving the eigenvalue problem $A\ket{\phi}=e^{i \phi}\ket{\phi}$, where the phase $\phi$ can be estimated to $n$-bit precision using an $n$-bit QFT \cite{Nielsen02}. These algorithms have been implemented in experiments using a semi-classical version of the QFT that consists of single-qubit rotations based on classical feed-forward and qubit recycling which reduces the required register size \cite{Monz15,Chiaverini05,Higgins07}. The coherent QFT, on the other hand, is reversible and can be concatenated within an algorithm sequence.

Here, we construct a coherent QFT on five qubits using all 10 modular CP gates and involving a total of 80 single- and two-qubit native gates. This circuit fully exploits the high connectivity of a trapped ion system and illustrates how it can be scaled to larger modules (Fig. \ref{fig:QFT}a). We apply the QFT in a period-finding protocol where we first prepare an input superposition state $\sum_{k=0}^{31}C_k\ket{k}$ such that the coefficients $\{C_k\}$ exhibit a periodic amplitude or phase modulation (see methods), which is followed by the QFT operation. The modulation periodicity then appears in the output state populations (Fig.\ref{fig:QFT}.b). 

We further examine the performance of the QFT in a phase estimation protocol where the eigenvalue $\phi$ is estimated to 5-bit precision. In this case the input state is prepared in the form $\frac{1}{\sqrt{32}}\otimes_{j=1}^5 ( \ket{0} + e^{-i 2^{j-1} \phi} \ket{1} )$, which exhibits a $\phi$-dependent phase modulation $C_k=\frac{1}{\sqrt{32}} e^{-i k \phi}$. We apply the QFT on this state to estimate $\phi$ by mapping its value onto populations of the output state, as shown in figure \ref{fig:QFT}c. This is repeated for several cases where $\phi$ is incremented in steps of $2\pi/64$ over the range $0$ to $2\pi$. Values of $\phi$ that are integer multiples of $2\pi/32$ result in the output state $\ket{32 \phi/2\pi}$. This is achieved with an average fidelity of $0.619(5)$. For non-integer values, the population is distributed between the nearest 5-bit approximate states \cite{Nielsen02}.

In our experiments, each algorithm fidelity is limited mainly by the native gate errors $(< 2\%)$, which propagate into the standard logic gate errors $(<5\%)$ (see supplementary materials). These errors are dominated by Raman beam imperfections and therefore can be reduced by mitigating Raman beam intensity noise \cite{Ballance16} and individual addressing crosstalk (see methods). Systematic shifts in the axes of the gate-rotations accumulate due to unequal Stark shifts across the qubits, which result in algorithmic errors that depend upon the circuit structure. This type of error can be easily eliminated by feeding forward known shifts to the rf of individual qubit control beams.

The algorithms presented here illustrate the computational flexibility provided by the ion trap quantum architecture. Within a single module, this system can be scaled to dozens of qubits by linearly increasing the number of rf controls, AOM- and PMT-channels at the hardware level. In software, the number of XX- and R-gate calibrations required to compile any logic gate scale as  $O(n^2)$. As more ions are added to the chain, the axial confinement must be weakened to maintain a linear crystal.  This will slow down the XX-gate duration roughly as $n^{1.7}$, but the crosstalk is not expected to get worse (see methods). Finally, implementing this architecture on multi-zone ion traps such as surface traps will provide further control over the connectivity of qubits though shuttling \cite{Kielpinski02} for scalable computation. This will also enable selective measurement of qubits that can be fed-forward classically to perform conditional operations in the module \cite{Chiaverini05} as required in fault-tolerant computing.


\bibliographystyle{plain}



\section{Acknowledgements}

We thank K. R. Brown, J. Kim, T. Choi, Z.-X. Gong, T. A. Manning, D. Maslov and C. Volin for useful discussions. This work is supported by the U.S. Army Research Office with funds from the IARPA MQCO Program, the Air Force Office of Scientific Research MURI on Quantum Measurement and Verification, and the National Science Foundation Physics Frontier Center at JQI.

\newpage
\clearpage
\onecolumngrid 



\renewcommand{\thetable}{\textbf{\arabic{table}}}  
\renewcommand{\thefigure}{} 
\renewcommand{\tablename}{\textbf{Table}}
\renewcommand{\figurename}{\textbf{Supplementary Figure 1}}

\newcommand{\FigureOneED}{
\begin{figure*}[t]
\includegraphics[]{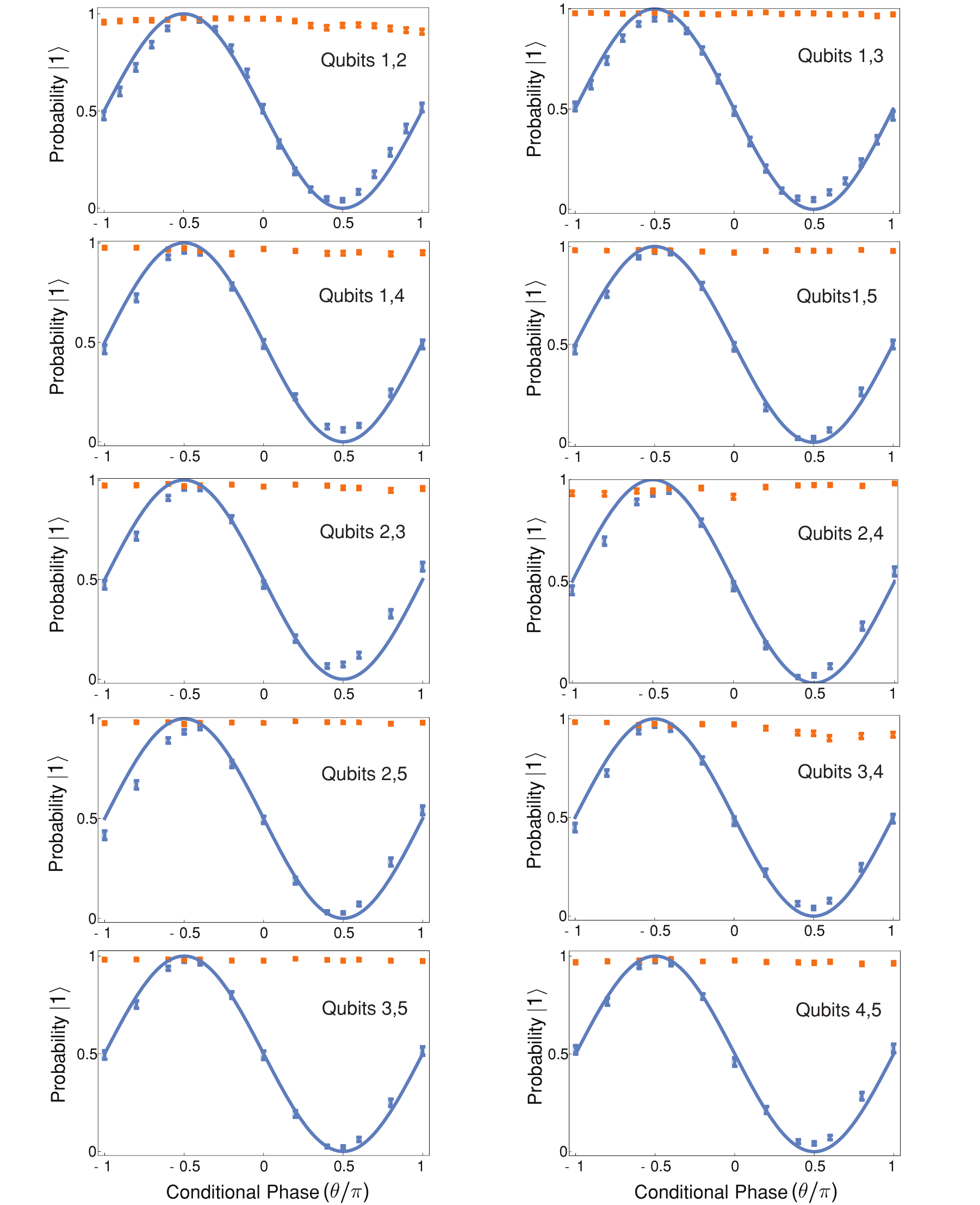}
\caption{\textbf{Controlled-Phase gate.} Controlled-Phase (CP) gate between control (red) and target (blue) qubit for different qubit-pairs.  The control qubit is prepared in the state $\ket{1}$ which remains unchanged during the gate.  Solid blue lines indicate theoretical probability of measuring the target qubit in $\ket{1}$ whereas the data points show experimental data. Error bars are statistical indicating a $95 \%$ confidence interval for 2,000 experimental repetitions.
}
\label{fig:CP5Ions}
\end{figure*}}


\section{Methods}
\textbf{Experimental techniques. }
We use a linear rf Paul trap made of four segmented blade electrodes driven at $23.83\:$MHz where the transverse secular frequency of the trap is actively stabilized \cite{Johnson16}. For measurement, state-dependent fluorescence is collected by a $0.38$ numerical aperture objective that images ions with $0.55\:\mu$m resolution. For a single qubit, single-shot detection fidelities for states $\ket{0}$ and $\ket{1}$ are $99.74(3)\: \% $ and $ 99.09(5)\: \% $, respectively. For $n=5$ qubits, detection is degraded by signal crosstalk between PMT channels, and the average single-shot fidelity is $95.3(2)\:\%$ for the $2^n$ states. For the population distributions measured in figures \ref{fig:DJBV} and \ref{fig:QFT} and the reported algorithm fidelities, multi-qubit detection is performed by signal-averaging the populations of all $2^n$ states over a few thousand experimental repetitions.  In this way, detection and crosstalk errors are removed by decomposing the measurements into the known detector array response of all $32$ possible qubit states. The individual addressing Raman beams are modulated using a multi-channel AOM \cite{Harris} and focused down to a beam waist of $\sim 1.5\:\mu$m at the ions. Addressing crosstalk between neighboring ions due to Raman beam spillover is $< 4\: \%$, which can be improved using higher resolution optics \cite{Crain14}.

As more ions are added to a chain, the ratio of axial-to-transverse confinement must be weakened to maintain a linear crystal $(\nu_z /\nu_x < 0.6 n^{-0.86})$ \cite{Schiffer93}. For constant transverse confinement, this means that the minimum ion spacing remains the same. However, this will slow the gates down. In our setup (for $n=5$) two-qubit XX-gates for any ion pair $\{i, j\}$ have a duration of $\tau_g = 235\: \mu$s, which depends on the spectral splitting of the transverse modes ($\tau_g \sim \nu_x/\nu_z^2 \sim n^{1.7}$). The XX-gate pulse shape is a 9-segment piecewise constant Rabi frequency modulation $\{\Omega_k\}_{ij}$ (where $1\le k\le 9$), which is implemented by modulating the global Raman beam. Optimized pulse shapes are calculated for each ion pair such that $\{\Omega_k\}_{ij}$ is within practical limits and the gate fidelity is maximized. The number of classical calculations to find the pulse shapes scales as $O(n^2)$. The XX-gates are calibrated by setting the product of the laser intensities on the two qubits such that $\chi_{i,j}=\pi/4$ \cite{Choi14,Zhu06, Solano99, Milburn00}. For CP gates that require other values of $\chi_{ij}$, we scale the laser intensity accordingly. Single-qubit rotations are calibrated by measuring the Rabi frequency $\Omega_i$ of individual qubits. Single-qubit native R-gates have a duration of $\sim 0.1\tau_g$. 

\textbf{Implementation of the Deutsch-Jozsa algorithm. } 
The Deutsch-Jozsa algorithm is implemented by starting with an equal superposition of all classical input states to the function  $f(x):\{0,1,\ldots,7\}\rightarrow \{0,1\}$. We prepare this by initializing all qubits to $\ket{0}$, followed by $R_{y}(\pi/2)$ rotations on the qubits in the control register $x=X_1X_2X_3$. Then we rotate the ancilla qubit $X_5$ using $R_{y}(-\pi/2)$. The resulting 5-qubit state is
\begin{equation*}
 \ket{\psi}_0= \frac{1}{\sqrt{8}}  \sum^{7}_{x=0}\ket{x}_{123}\otimes\ket{0}_{4}\otimes\frac{\ket{0}_5-\ket{1}_5}{\sqrt{2}}
\end{equation*}
where $x$ is the decimal representation of qubits $X_1X_2X_3$. Then we apply the function on the input superposition state such that the value is written to $X_4$. The resulting state is
\begin{equation*}
 \ket{\psi}_1=\frac{1}{\sqrt{8}} \sum^{7}_{x=0}\ket{x}_{123} \ket{f(x)}_{4} \otimes\frac{\ket{0}_5-\ket{1}_5}{\sqrt{2}}
\end{equation*}
This is followed by a CNOT between the function register $X_4$ and the ancilla $X_5$ which provides a phase ``kick-back'' to produce the state
\begin{equation*}
 \ket{\psi}_2=\frac{1}{\sqrt{8}} \sum^{7}_{x=0}(-1)^{f(x)}\ket{x}_{123} \ket{f(x)}_4 \otimes\frac{\ket{0}_5-\ket{1}_5}{\sqrt{2}}
\end{equation*}
This is followed by a single-qubit rotation $R_{y}(\pi/2)$ on all qubits. Then we measure the first four qubits to reach the solution and ignore the ancilla qubit since it is not entangled with the other qubits. The state of qubits $X_1X_2X_3X_4$ before measurement can be written as

\begin{align}\label{eqn1}
\ket{\psi}_3 & =\frac{1}{8} \sum^{7}_{y=0}\sum^{7}_{x=0}(-1)^{f(x)}(-1)^{\bar{y}\cdot x}\ket{\bar{y}}_{123}\otimes \frac{\ket{1}_4+(-1)^{f(x)}\ket{0}_4}{\sqrt{2}} \\ 
\nonumber &=C_{0000}\ket{0000}+C_{0001}\ket{0001}+ \ldots \\
\nonumber & \ldots +C_{1110}\ket{1110}+C_{1111}\ket{1111}
\end{align}
where $\bar{y}$ is the bit-wise inversion of $y$. If $f(x)=a$ is a constant function (with $ a=\{0,1\}$), the coefficients of the basis states $\ket{1110}$ and $\ket{1111}$ are
\begin{align*}
 C_{1110}&=\frac{1}{8\sqrt{2}} (-1)^{a} \sum^{7}_{x=0} (-1)^{000\cdot x} = \frac{(-1)^{a}}{\sqrt{2}}\\
 C_{1111}&=\frac{1}{8\sqrt{2}} \sum^{7}_{x=0} (-1)^{000\cdot x} = \frac{1}{\sqrt{2}}
\end{align*}
If $f(x)$ is a balanced function, then the coefficients are 
\begin{align*}
 C_{1110} &=\frac{1}{8\sqrt{2}} \sum^{7}_{x=0} (-1)^{000\cdot x}(-1)^{f(x)}(-1)^{f(x)} = \frac{1}{\sqrt{2}}
\\
 C_{1111}&=\frac{1}{8\sqrt{2}} \sum^{7}_{x=0} (-1)^{000\cdot x}(-1)^{f(x)} = 0
\end{align*}
Here we use the property that $f(x)=0$ for exactly half of the values of $x$ and 1 for the rest. Conditioned upon $X_4=1$, there is unit probability of measuring $X_1,X_2,X_3=111$ for a constant function and $0$ probability of measuring the same outcome when the function is balanced. In equation \ref{eqn1}, note that the probability of measuring $X_4=1$ is $0.5$ irrespective of the number of qubits in the input (control) register of the function.

\textbf{Native single- and two-qubit rotations. }
Native single-qubit operations $R_{\phi}(\theta)$ are rotations of the Bloch vector by an angle $\theta$ about an axis on the equator of the Bloch sphere, where $\phi$ is the angle between this rotation axis and the X-axis. The single-qubit operator is
\begin{equation*}
 R_{\phi}(\theta)=\begin{bmatrix}
\text{cos}(\frac{\theta}{2})& -i \text{sin}(\frac{\theta}{2})e^{-i\phi} \\
-i \text{sin}(\frac{\theta}{2})e^{i\phi} &  \text{cos}(\frac{\theta}{2})
\end{bmatrix}
\end{equation*}
The standard X and Y rotations used in the composite gates are simply $R_x(\theta)=R_{0}(\theta)$ and $R_y(\theta)=R_{\pi/2}(\theta)$.

Native two-qubit XX-gates are performed by invoking a $\sigma_x\sigma_x$-Ising interaction between qubits $i$ and $j$, which is mediated though the coupling of the qubits to the collective transverse motional modes of the ion chain. The resulting two-qubit entangling rotation $XX(\chi_{ij})$ depends on the geometric phase $\chi_{ij}$, which is the integrated Ising interaction and can be varied by changing the Raman beam intensity. The sign of the geometric phase $\alpha_{ij}=\text{sgn}(\chi_{ij})$ depends on how ions $i$ and $j$ couple to the common transverse motional modes. The XX-gate operator is
\begin{align*}
 XX&(\chi_{ij})=\\
&\begin{bmatrix}
\text{cos}(\chi_{ij}) & 0 & 0 & -i\: \text{sin}(\chi_{ij})\\
0 & \text{cos}(\chi_{ij}) & -i\: \text{sin}(\chi_{ij}) & 0\\
0 & -i\: \text{sin}(\chi_{ij}) & \text{cos}(\chi_{ij}) & 0\\
-i\: \text{sin}(\chi_{ij}) & 0 & 0 & \text{cos}(\chi_{ij})\\
\end{bmatrix}
\end{align*}
In this experiment, $\alpha_{12},\alpha_{45},\alpha_{14},\alpha_{25},\alpha_{35},\alpha_{23},\alpha_{34}=+1$ and $\alpha_{15},\alpha_{25},\alpha_{13}=-1$.

\textbf{Composite gate fidelity. }
Controlled-NOT (CNOT) gates are performed between all ion pairs and characterized in the following way. We perform the CNOT gate on all four classical input states $\ket{00}, \ket{01}, \ket{10}, \ket{11}$ and measure the fidelity from the population of the desired output state. The average fidelity of a CNOT on each ion pair is shown in Table \ref{CNOT}.

Controlled-Phase (CP) gates are performed between all ion pairs and characterized by using a sequence of gates. We first initialize the qubits in the state $\frac{1}{\sqrt{2}}\ket{1}(\ket{0}+\ket{1})$, where the first qubit is the control qubit and the second qubit is the target qubit. This is followed by a conditional phase gate $\text{CP}(\theta)$ that creates the state $\frac{1}{\sqrt{2}}\ket{1}(\ket{0} +e^{i \theta}\ket{1})$. A final rotation $R_x (\frac{\pi}{2}) $ on the target qubit projects the conditional phase $\theta$ onto the population of the target qubit as $P(\ket{1})= \frac{1}{2}(1-\text{sin}\theta)$. This is shown in Supplementary Figure {\bf 1}. 

We measure the fidelity of the CP gates at conditional phases $\theta = \pm \frac{\pi}{2}$, which correspond to the maximum and minimum values of $\theta$, respectively, which are used in a coherent QFT or QFT$^{-1}$. At these values of $\theta$, where the geometric phase $\chi_{ij}=\frac{\pi}{4}$, the XX-gates are most sensitive to laser intensity fluctuations, which leads to maximum errors. This is evident from the data shown in Supplementary Figure {\bf 1}, where a maximum deviation of the analysis qubit from the ideal output state occurs at $\pm \frac{\pi}{2}$. Therefore, the fidelity measure at these values is a lower bound on the CP gate fidelity. The fidelity is obtained by  measuring the populations in the $\ket{10}$ and $\ket{11}$ states for $\theta = +\frac{\pi}{2}$ and $\theta = -\frac{\pi}{2}$, respectively. Table \ref{CP} shows the fidelities of all CP gates.

\textbf{QFT state preparation. }
For the period-finding experiment, an amplitude or phase modulation is created in the coefficients ${C_k}$ of the input state $\sum_{k=0}^{31} C_k \ket{k}$ using individual single-qubit rotations. Table \ref{QFT} shows the input states for various measured periodicities.


\begin{table}[h!]
\caption{Controlled-NOT gate fidelities for 5 qubits} \label{CNOT} 
	\begin{center}
 	\begin{tabular}{| l | c || l | c |}
	\hline
	Ion pair & Fidelity (\%) & Ion pair & Fidelity(\%)\\
	\hline
  	1,2 & 96.4(6) & 2,4 &98.5(7)\\
 	\hline
 	 1,3 & 97.6(7) & 2,5 & 96.8(7)\\
 	\hline
	  1,4 & 95.9(7) & 3,4 & 96.6(5)\\
	\hline
	  1,5 & 97.9(5) & 3,5 & 97.6(6)\\
	\hline
	  2,3 & 95.6(6) & 4,5 & 97.2(5)\\
	\hline
	\end{tabular}
	\end{center}
\end{table}

\begin{table}[h!]
\caption {Controlled-Phase gate fidelities for 5 qubits} \label{CP} 
	\begin{center}
	\begin{tabular}{| l |c |c |}
	\hline
	Ion pair & $\theta = \frac{\pi}{2}$, fidelity (\%) & $\theta = -\frac{\pi}{2}$, fidelity (\%)\\
	\hline
  	1,2 & 91.1(6) & 96.1(4)\\
 	\hline
 	 1,3 & 93.6(5) & 93.3(6)\\
 	\hline
	  1,4 & 91.6(6) & 93.3(6)\\
	\hline
	  1,5 & 95.9(4) & 95.3(3) \\
	\hline
	  2,3 & 90.7(6) & 93.2(5)\\
	\hline
	  2,4 &  94.2(5)& 90.8(6) \\
	  \hline
	  2,5 & 95.8(4) & 91.7(6) \\
	  \hline
	  3,4 & 91.0(6) & 94.7(5) \\
	  \hline
	  3,5 & 96.0(4) & 96.0(4) \\
	  \hline
	  4,5 & 93.5(6)  & 95.8(4) \\
	 \hline
	\end{tabular}
	\end{center}
\end{table}

\begin{table}[h!]
\caption{Input states in QFT-period finding} \label{QFT} 
	\begin{center}
	\begin{tabular}{| l |c |}
	\hline
	Input state & Period \\
	\hline
	$\frac{1}{\sqrt{32}}(\ket{0}+\ket{1})(\ket{0}+\ket{1})(\ket{0}+\ket{1})(\ket{0}+\ket{1})(\ket{0}+\ket{1})$& 1 \\
	\hline
  	$\frac{1}{\sqrt{32}}(\ket{0}+\ket{1})(\ket{0}+\ket{1})(\ket{0}+\ket{1})(\ket{0}+e^{i6.2\pi/16}\ket{1})(\ket{0}+i\ket{1})$& 3\\
 	\hline
 	 $\frac{1}{\sqrt{8}}(\ket{0}+\ket{1})(\ket{0}+\ket{1})(\ket{0}+\ket{1})\ket{11}$ & 4\\
 	\hline
	 $\frac{1}{2}(\ket{0}+\ket{1})(\ket{0}+\ket{1})\ket{111}$ & 8\\
	\hline
	 $\frac{1}{\sqrt{2}}(\ket{0}+\ket{1})\ket{1111}$ & 16 \\
	\hline
	 $\ket{11111}$ & 32\\
	\hline
	\end{tabular}
	\end{center}
\end{table}


\FigureOneED


\end{document}